\documentclass[reprint,amsmath,amssymb,aps,pre]{revtex4-2}
\usepackage{upgreek}

\usepackage{graphicx}% Include figure files
\usepackage{dcolumn}% Align table columns on decimal point
\usepackage{bm}% bold math
\begin{document}

\preprint{APS/123-QED}

\title{Relaxation pathways in X-ray Free Electron Laser heated Iron}% Force line breaks with \\

\author{L. Ansia}%
\email{lucas.ansia.fernandez@tecnico.ulisboa.pt}
\affiliation{%
GoLP/Instituto de Plasmas e Fusão Nuclear, Universidade de Lisboa, 1049-001, Portugal}
\affiliation{
 Instituto de Fusión Nuclear, Universidad Politécnica de Madrid, José Gutiérrez Abascal 2, 28006, Spain.
}%

\author{G. O. Williams and M. Fajardo}%
\affiliation{%
GoLP/Instituto de Plasmas e Fusão Nuclear, Universidade de Lisboa, 1049-001, Portugal.
}%

\author{P. Velarde}
\affiliation{
 Instituto de Fusión Nuclear, Universidad Politécnica de Madrid, José Gutiérrez Abascal 2, 28006, Spain.
}%

\begin{abstract}
Non-thermal photo-ionized plasmas are now established in the laboratory, and require models that treat the atomic processes and electron distribution self-consistently. We investigate the effects of inelastic thermalization in iron under intense X-ray irradiation using the atomic model BigBarT, suited for the self-consistent evolution of the electron continuum, including degeneracy effects. Our study focuses particularly on collisional $M$-shell ionization, which we identify as the dominant relaxation process of the non-thermal electrons. We show that $M$-shell satellite intensities are sensitive to non-thermal ionization, providing a potential method to refine collisional cross sections that are otherwise difficult to compute due to their proximity to the continuum and the associated plasma screening effects. 
\end{abstract}

\maketitle

%\tableofcontents

\section{\label{sec:level1}Introduction\protect\\}

Non-equilibrium dynamics are central to understanding complex many-body behaviours across various disciplines. In high-energy-density plasmas, this topic is increasingly important due to its relevance in inertial fusion, where hot electrons can be detrimental to achieving optimal compression and energy gain. \cite{Fusion1, Fusion2, Fusion4}.

X-ray free-electron lasers (XFELs) create a highly non-equilibrium system among photons, electrons, and ions. The processes that generate these hot electron distributions are generally well understood, linear, and can be modelled with reasonable accuracy. However, the relaxation pathways depend on collisional cross sections, which are often approximated or derived from gas-phase experiments. This becomes increasingly challenging for higher quantum numbers, especially as they approach the continuum. 

In the past two decades, XFELs have made significant advancements, achieving peak intensities in the XUV and X-ray regions, previously only possible in the optical and infrared ranges. They allow for the creation of samples with greater uniformity and well-defined properties such as temperature and density \cite{XFEL1, XFEL2}, while achieving peak intensities exceeding 10$^{20}$ Wcm$^{-2}$. This technology has enabled unprecedented characterization of plasmas, critical in astrophysics for studying conditions in planetary interiors \cite{INTERIOR1, INTERIOR2}. 

Collisional Radiative Models (CRM) are particularly useful simulations tool, as self-emission serve as the primary diagnostic for XFEL driven plasma \cite{CRM1}. These models have been extensively employed to simulate XFEL experiments under various conditions \cite{CRM2, CRM3}. They allow to recreate, based in the fluorescence spectra, the temporal evolution of the plasma, focusing on the atomic kinetics. Particularly important to this work are the results of Q. Y. van den Berg et al., in which magnesium was resonantly pumped at the 1s-2p transition to measure the CI cross-section by comparison with several existing models \cite{CRM4}. 

Several efforts have been made to couple collisional-radiative models (CRM) with non-equilibrium dynamics. One of the earliest applications of the Fokker-Planck approach for self-consistently evolving continuum electrons was introduced by J. Bretagne et al., who studied electron-beam-generated argon plasmas \cite{OG}. This approach was later extended to X-ray-produced plasmas by J. Abdallah et al., which coupled atomic rate equations with the self-consistent evolution of distributions \cite{OG1, OG2}. These developments enabled one of the first studies on XFEL-driven electron dynamics \cite{OG3}. Building on this foundation, de la Varga and colleagues developed the BigBarT code and parallelly investigate the effects of non-Maxwellian plasmas in X-ray-pumped neon \cite{BIGBART}, later incorporating degeneracy to better more dense plasmas \cite{BIGBART2}.  

Recently, similar approaches have been applied to solid-density plasmas. For instance, CCFLY, an extension of the SCFLY code, has been used to study non-thermal emission signatures in aluminum and magnesium plasmas \cite{NT1}. Additionally, it was also employed to propose a method for refining the Coulomb logarithm \cite{NT2}. Cheng Gao and collaborators further examined the effects of non-thermal dynamics on XFEL transmission in solid aluminum \cite{CRM6}.

Despite these advances, studies on higher-\( Z \) materials remain limited due to their computational complexity. B. Ziaja et al. developed a kinetic Boltzmann code designed for modelling X-ray-created WDM by reducing the number of atomic paths in the CRM system \cite{BOLT1, BOLT2, BOLT3}. This approach has been recently used to characterize XFEL transmission in copper at intensities up to \( 10^{17}\,\text{Wcm}^{-2} \) \cite{BOLT4}.  Hai P. Le et al. looked at the interplay between atomic kinetics and non-thermal electron, focusing on inverse bremsstrahlung heating and non-local transport effects, in materials up to molybdenum \cite{SCOTT}.  

For this work, we adapted the BigBarT code to simulate arbitrary materials, with a particular focus on the first transition metals. It models the self-consistent evolution of the non-thermal part of the electronic continuum. The code also accounts for degeneracy in both the electron distribution and rate calculations which has been shown to be  important in solid density density \cite{CRM5, CRM7}. 

We investigate the impact of non-thermal electron distributions on the evolution of XFEL heated, solid-density iron. Additionally, we explore the effect of modifying outer collisional ionization cross-sections. This is particularly interesting because precise description of near-continuum wave functions and many-body effects remains a significant challenge for determining cross-section values. Inelastic collisions, as the primary thermalization mechanism for high energy electrons could leave observable signatures that help validate existing models. Iron was chosen for our simulations due to its astrophysical importance and the lack of studies addressing the impact of non-thermal electron distributions on its emitted spectra, thereby extending prior work on lower-\( Z \) materials \cite{NT1}. 

The paper is organized as follows. In section \ref{sec2} we describe BigBarT, both the computational framework and the simplifications made to make the problem tractable. In section \ref{sec3} we devote a section to clarify the evolution of a plasma when subjected to X-ray radiation, emphasising on the effects of non-thermal distributions and Fermi-Dirac statistics. Lastly, in section \ref{sec4} non-equilibrium effects on the self-emission are discussed.

% (**GW is $\varepsilon_1$ needed? why not just $\varepsilon$ ? and there is $\varepsilon$ in the integration limits ** )

\section{Simulation approach}\label{sec2}

Simulations have been carried out using the CRM BigBarT \cite{BIGBART}. It couples the atomic rate equations with a degenerate Fokker-Planck approach to self-consistently evolve the ion and electron distributions. In this fashion, non thermodynamic equilibrium is assumed in both systems. Ionic populations are updated following the usual rate equation, 

\begin{equation}
    \frac{d n_i}{d t} = \sum_{j \neq i}^{m} n_j R_{ji} - n_i \sum_{j \neq i}^{m} R_{ij},
\end{equation}

Where \( n_i \) is the population of each state, and \( R_{ij} \) and \( R_{ji} \) represent the total rates between complex \( i \) and \( j \) respectively. 

The processes include photo-ionization (PI), collisional ionization (CI), Auger effect (AU), collisional excitation (CE) and spontaneous emission (SE), along with their inverse processes. The cross sections for PI, SE, and AU are obtained in relativistic detail using the configuration average mode (i.e., UTA) of the Flexible Atomic Code \cite{FAC}. CE is calculated using the plane-wave Born (PWB) approximation with the JJATOM model \cite{CE}, which allows for efficient computation at the relativistic level and compares well with those obtained by FAC and HULLAC in the distorted wave approximation. 

\begin{figure*}
\includegraphics{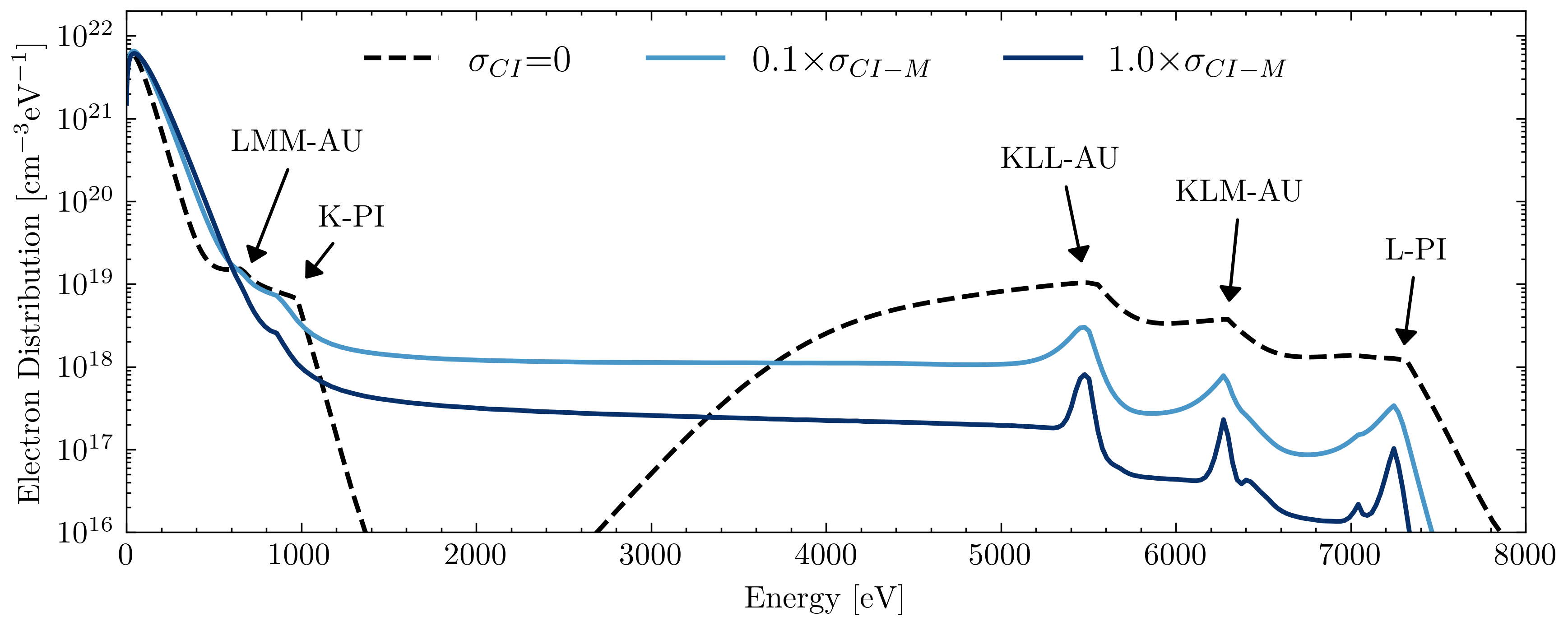}% Here is how to import EPS art
\caption{\label{fig:wide} Electron distributions at the pulse peak (60 fs) for varying collisional ionization cross sections of the M-shell. The dotted line represents the scenario without collisional ionization under the same conditions. Key features of photoionization (PI) and Auger (AU) processes are highlighted. For PI, the labels indicate the photoionized shell, while for AU, the first two labels correspond to bound-bound transitions, and the last one designates the ionized level. Conditions for all simulations are 30 fs FWHM pulse at $1 \times 10^{18} \, \text{Wcm}^{-2}$, and an incoming photon energy of 8100 eV.}
\label{Distribution_comp}
\end{figure*}

The CI differential cross section is obtained using the coulomb born exchange fitting from C. Fontes and colleagues \cite{CI1, CI2}. Non-relativistic cross sections for any type of ion can be obtained in significantly less time and can include IPD scaling convenient for dense scenarios. Moreover, this fitting has been successfully applied to warm dense matter (WDM) experiments before \cite{CRM4}. All inverse processes are obtained by means of the microscopic reversibility relations. 

To reduce computational effort, the super-configuration approach has been adopted, where complex states are only distinguished by their principal quantum number. In this paper, we follow the usual $KLM$ nomenclature, where each number represents the number of electrons in the $K$, $L$, or $M$ shell, respectively (e.g., an arbitrary complex with the first two shells completely filled and an extra electron in the third level, will be formed by the relativistic configurations \(3s_{1/2}\), \(3p_{1/2}\), \(3p_{3/2}\), \(3d_{3/2}\) and \(3d_{5/2}\), and referred to as 281). Transition rates are thus grouped, forming initial and final super-configurations, taking into account the statistical weight of each relativistic level.

Rates are calculated as functions of the atomic cross section and the electron distribution function. To capture the effect of degeneracy, Pauli blocking is included.   Due to its importance in this work we explicitly write here the used expression for CI and three body recombination (3B) process. Given an incident electron of energy $\varepsilon_1$, ionization potential $I$ and two outgoing electrons of energy $\varepsilon_{1'}$ and $\varepsilon_{2'}$ such that $\varepsilon_1 = \varepsilon_{1'} + \varepsilon_{2'} - I$, 

\begin{widetext}
\begin{equation}
    R_{CI} = \int \int_{0}^{\frac{\varepsilon_1-I}{2}} \sqrt{\frac{2 \varepsilon_1}{m_e}} g(\varepsilon_{1}) \tilde{f}(\varepsilon_{1}) \sigma_{ci}(\varepsilon_1, \varepsilon_{1'}) 
    P(\varepsilon_{1'}) P(\varepsilon_{1}-\varepsilon_{1'}-I) d\varepsilon_{1'}
    d\varepsilon_{1}
    \label{CI}
\end{equation}

\begin{equation}
    R_{3B} = \frac{g_i}{g_j} \frac{h^3}{8 \pi m_e^2} \int_I^{\infty} \varepsilon_1 P(\varepsilon_{1}) \left(\int_{0}^{\frac{\varepsilon_1-I}{2}} \frac{g(\varepsilon_{1'}) \tilde{f}(\varepsilon_{1'})g(\varepsilon_{1}-\varepsilon_{1'}-I)\tilde{f}(\varepsilon_{1}-\varepsilon_{1'}-I)}{\sqrt{\varepsilon_{1'}(\varepsilon_{1}-\varepsilon_{1'}-I)}} \sigma_{ci}(\varepsilon_1, \varepsilon_{1'}) d \varepsilon_{1'}  \right) d \varepsilon_{1}
    \label{3B}
\end{equation}
\end{widetext}

 Where, $g_i$ and $g_j$ are the statistical levels of the ionizing and ionized levels, $h$ is the planck constant and $m_e$ is the electron mass. \( g(\varepsilon) \propto \sqrt{E} \) represents the density of states, which determines the number of available levels at a given energy. The function \( \tilde{f}(\varepsilon) \) is the occupation factor, which varies between 0 (no electrons occupying the state) and 1 (full occupation of the state). The total electron density is given by the integral
$n_e = \int f(\varepsilon) d\varepsilon = \int g(\varepsilon) \tilde{f}(\varepsilon) \, d\varepsilon$. Finally, \( P(E) = 1 - \tilde{f}(\varepsilon) \) represents the Pauli blocking factor, which describes the probability that a state is unoccupied and thus available for transitions. The 3B rate employs the CI differential cross section since the Fowler relation have been inserted \cite{FOWLER}. 

The electron distribution evolves according to the Fokker-Planck equation \cite{FPT1, FPT2, FP2},

\begin{equation}
\begin{aligned}
\frac{1}{\Gamma} \frac{\partial f(\mathbf{v},t)}{\partial t} =  & - \nabla_{\mathbf{v}} \cdot \left[ \mathbb{H}(\mathbf{v},t) f(\mathbf{v}) \right] \\
& + \frac{1}{2} \nabla_{\mathbf{v}} \nabla_{\mathbf{v}} : \left[ \mathbb{G}(\mathbf{v},t) f(\mathbf{v}) \right] \\
& + \left( \frac{\partial f(\mathbf{v})}{\partial t} \right)_{\text{I}} \\
& + S(t),
\end{aligned}
\end{equation}

\noindent which is coupled to the atomic system via source terms, \( S \), that include photo-ionization and Auger effects, as well as inelastic terms \( ( \tfrac{\partial f(\mathbf{v})}{\partial t})_{\text{I}} \), which account for collisional ionization and collisional excitation, along with their inverse processes. 

 $\Gamma$ is defined as,   \begin{equation}
\Gamma = \frac{n_0 e^4 \ln \Lambda}{4 \pi \epsilon_0^2 m_e^2},
\end{equation}
where \(n_0\) is the electron density, \(e\) is the electron charge, \(m_e\) is the electron mass, and \(\ln \Lambda\) is the Coulomb logarithm. \( \mathbb{H}(\mathbf{v},t) \) and \( \mathbb{G}(\mathbf{v},t) \) are the Rosseblunt potentials \cite{FP2} associated with friction and diffusion, defined as:

\begin{equation}
\mathbb{H}(\mathbf{v},t) = 2 \cdot  \nabla_{\mathbf{v}} \int d\mathbf{v'} \frac{f(\mathbf{v'},t)}{|\mathbf{v} - \mathbf{v'}|},
\end{equation}

\begin{equation}
\mathbb{G}(\mathbf{v},t) = \nabla_{\mathbf{v}} \nabla_{\mathbf{v}} \int d\mathbf{v'} |\mathbf{v} - \mathbf{v'}| f(\mathbf{v'},t),
\end{equation}

\noindent here $\boldsymbol{\nabla_{\mathbf{v}} \nabla_{\mathbf{v}}}$ and $\boldsymbol{\nabla_{\mathbf{v}} \nabla_{\mathbf{v}}} :$ are the tensor generalization of the gradient and divergence, respectively \cite{FPT1}.

The numerical solution is carried out using the scheme presented by Bobylev et al. \cite{FP3}, which was later applied by Tzoufras and colleagues to study high-energy-density physics \cite{FP1}. Spatial dimensions are not considered, and thus only the isotropic component is solved. This scheme has been modified to account for Fermi-Dirac statistics \cite{FP_DEGEN, BIGBART2} and ensures particle and energy conservation. Both systems are evolved at each time step as an initial value problem using the CVODE solver from the SUNDIALS package \cite{SUN1, SUN2}.  

Ionization potential depression (IPD) is calculated using the Ecker-Kröll (EK) model \cite{EK} based on the initial conditions and is held constant throughout the simulation. This simplification is adopted as the correct way to treat the self-consistent evolution of the continuum with a dynamically varying IPD is still an open question.

\begin{figure}
\includegraphics[scale=0.85]{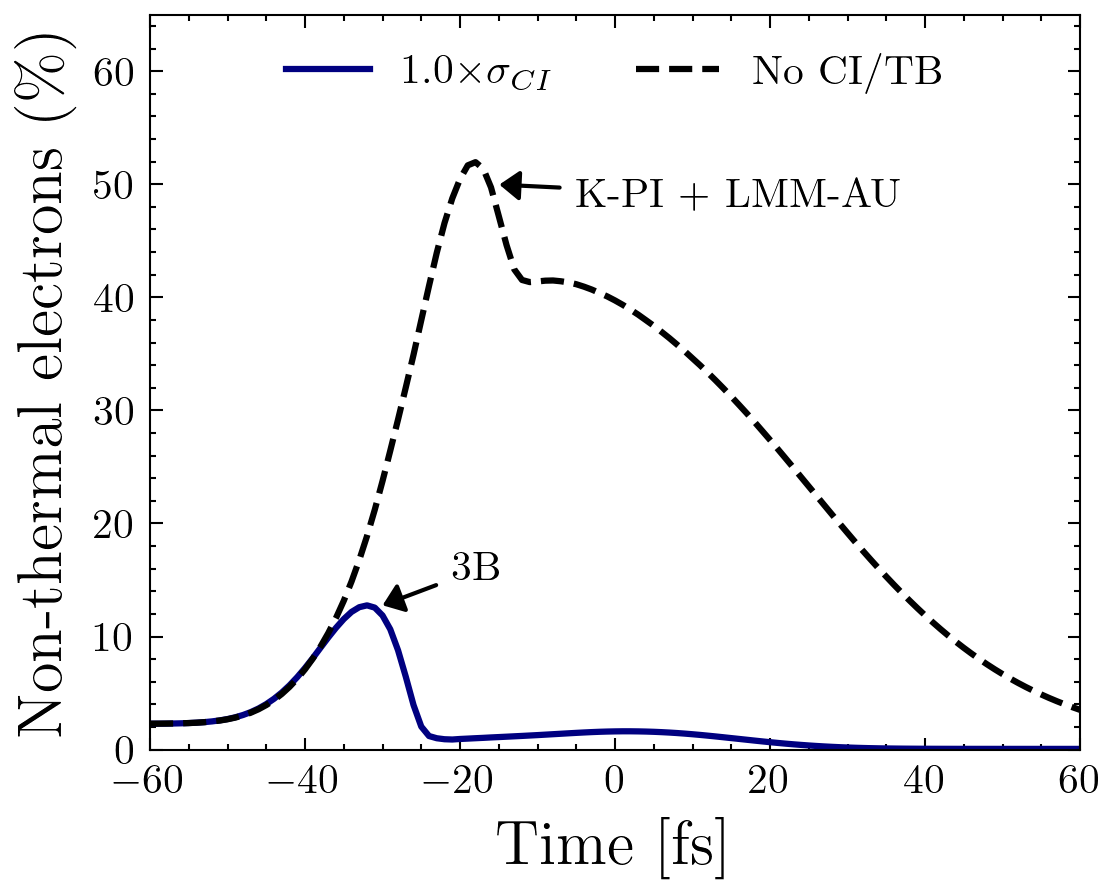}% Here is how to import EPS art
\caption{\label{delta} Percentage of electrons out of equilibrium ($\delta$) as a function of time for a fully inelastic simulation compared to one without collisional ionization and three-body recombination. Conditions for both simulations are 30 fs FWHM pulse at $1 \times 10^{18} \, \text{Wcm}^{-2}$, and an incoming photon energy of 8100 eV.
}
\end{figure}

The Coulomb logarithm, a crucial factor for accurate electron-electron collision simulations, faces similar challenges. Various models have been proposed, many of which exhibit divergent behavior at lower temperatures. We adopt the formulation by Shaffer and Starret \cite{Coulomb} , which, in the low-temperature limit, tends to the functional form:

\begin{equation}
    \ln \Lambda_{ee} = \frac{5}{4} \text{erf} \left[ \left( \frac{2T}{3T_F} \right)^3 \right]
\end{equation}

As stated by the authors, it does not hold any special physical significance but shows substantially improved low-temperature behaviour. We explicitly switch for the classical coulomb logarithm when the plasma conditions are no longer degenerate, $\Theta = \frac{\mu}{ k_{\beta} T_e} < 0$. Even so, as will be explained in the next section this factor will not influence the overall conclusions.

All the simulations presented here are for iron. The initial conditions are determined by solving the degenerate Saha equation, which yields an initial configuration of 288, with the $3p$ electrons being the last bound levels.

\section{Non-thermal evolution}\label{sec3}

The interaction between X-rays and solid matter primarily involves photo-ionization, which generate a stream of non-thermal electrons while creating inner shell holes. These holes filled either by spontaneous emission or through the Auger effect, the latter acting as an additional source. Once generated, electrons thermalize through two main mechanisms: elastic collisions between themselves and inelastic collisions with ions. Among them, CI is the most significant mechanism affecting the overall electron distribution.

\begin{figure}
\includegraphics[scale=0.85]{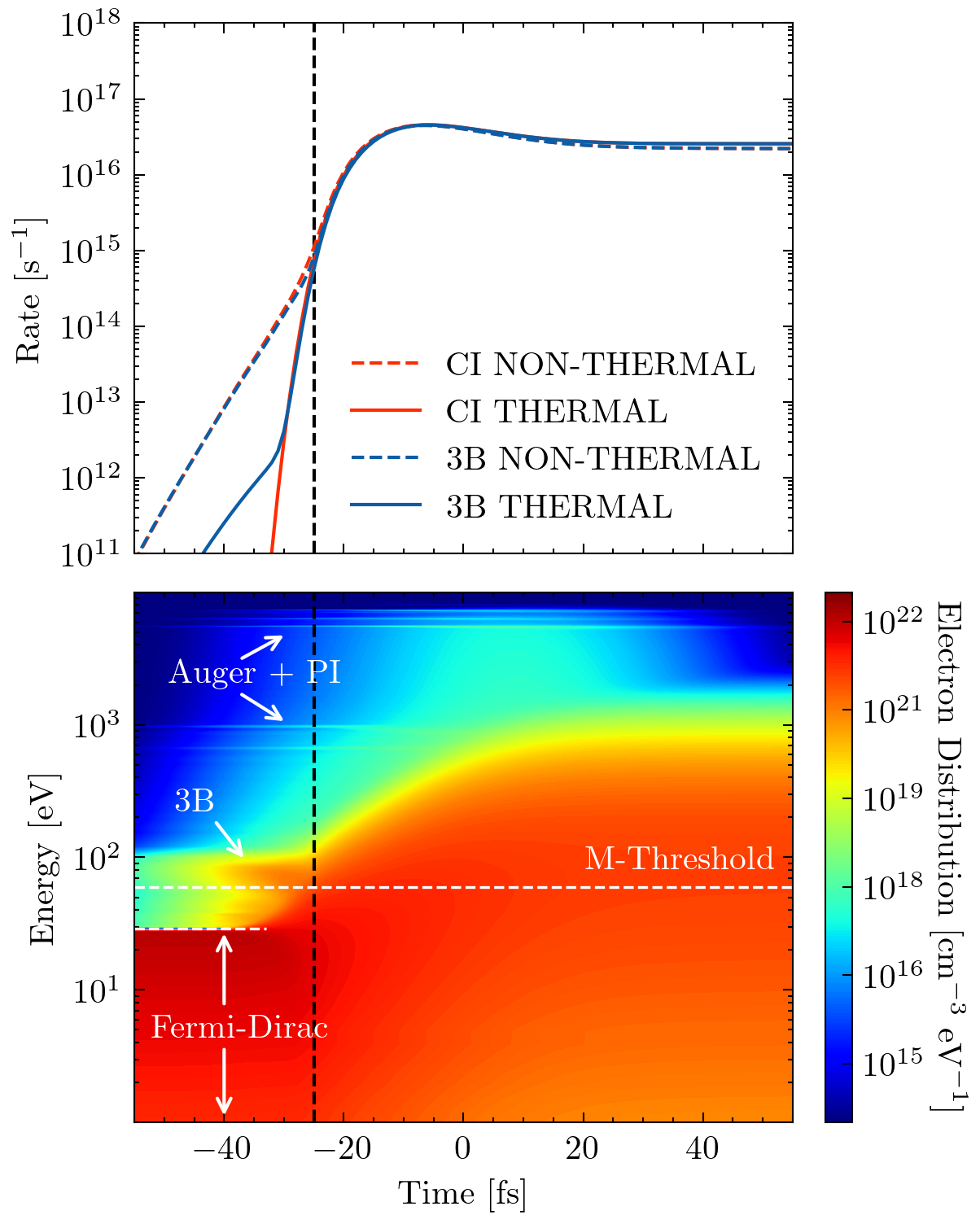}% Here is how to import EPS art
\caption{\label{Rates} Temporal evolution of the total collisional ionization (CI) and three-body recombination (3B) rates for both thermal and non-thermal simulations (top), along with the electron distribution for the non-thermal case (bottom), both plotted as functions of time. The M-threshold represents the minimum energy required to ionize the outermost electrons. The black dotted line indicates the point where the thermal (TH) and non-thermal (NT) rates converge and its intersection with the M-threshold. Conditions for both simulations are 30 fs FWHM pulse at $1 \times 10^{18} \, \text{Wcm}^{-2}$, and an incoming photon energy of 8100 eV. In the time axis, 0 correspond to the peak of the pulse.
}
\end{figure}

Various snapshots of the electron distribution at peak pulse for different values of the $M$-shell CI cross sections, $\sigma_{CI-M}$, including one with suppressed CI, are shown in Figure \ref{Distribution_comp} for a 30 fs FWHM pulse at $1 \times 10^{18} \, \text{Wcm}^{-2}$, and an incoming photon energy of 8100 eV. Despite differences in the overall shape, the main features are consistent across all simulations. Since the $K$-edge of Fe sits around 7.1 keV, the main photo-peak will be at 1000 eV. Peaks around 800 eV correspond to the Auger peak for the 278 \(\rightarrow\) 286 transition. Higher energy features correspond $L$ and $M$ photo-peaks and other combination of Auger corresponding to $L$ \(\rightarrow\) $K$ shells decay. Smaller signatures arise from collisional excitation/de-excitation of main features.

\begin{figure}
    \centering
    \includegraphics[scale=0.7]{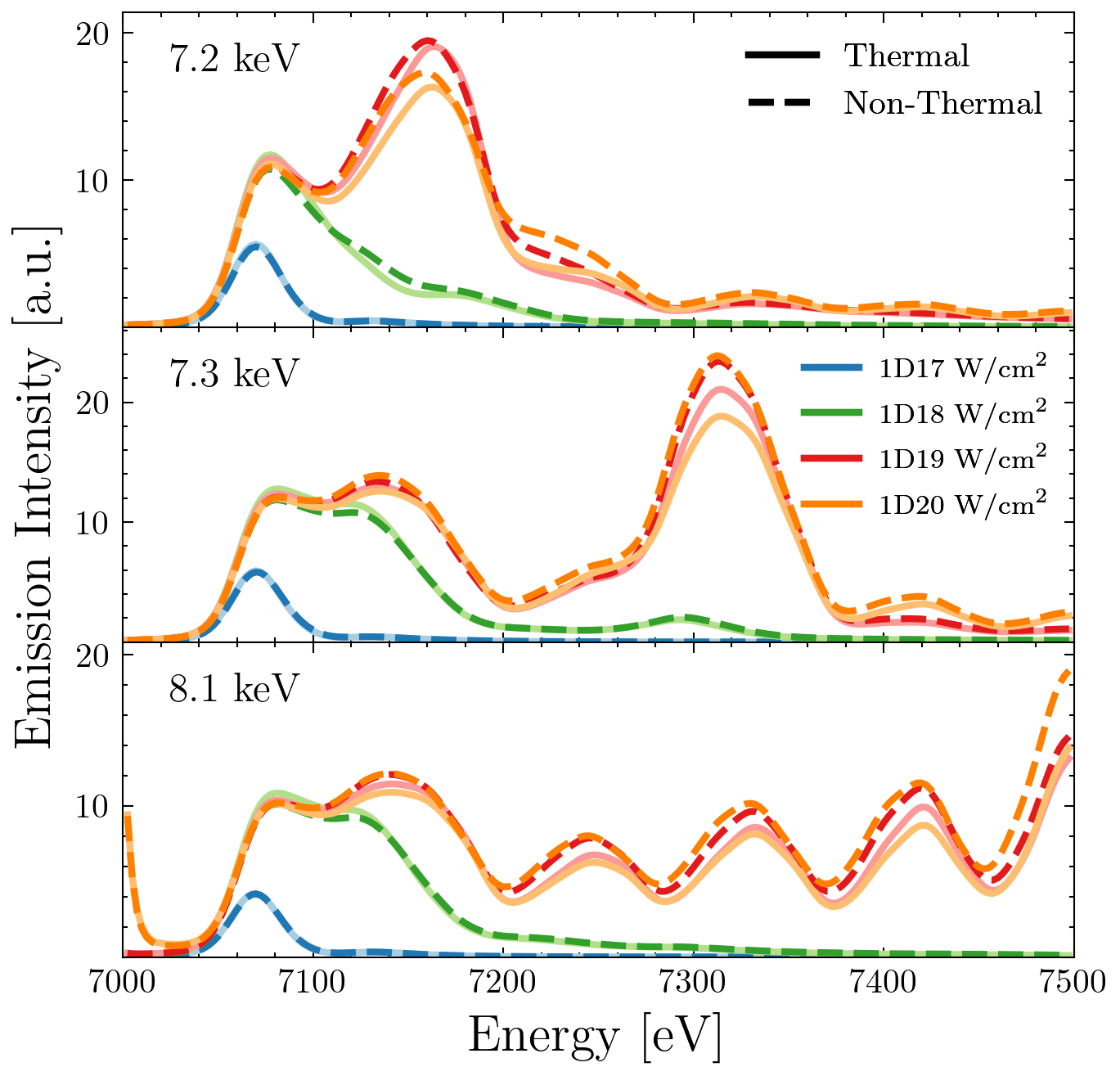}
    \caption{\label{S3R} Comparison of thermal and non-thermal spectra at varying intensities for the three pump energies analysed in this study. Results are shown for a 30 fs FWHM pulse. Spectra is focused in the first 4 $K_{\beta}$ $L$-satellites. The top plot shows only the first $L$-satellite. The middle plot displays the main \(K_{\beta}\) line along with the \(L6\) resonant line. The bottom plot highlights the first four satellites: $L8 \sim 7130$, $L7 \sim 7240$, $L6 \sim 7130$, and $L5 \sim 7420$, all of which are off-resonant.}
\end{figure}

The importance of inelastic collisions is clearly illustrated by the effect of their absence (black dotted line in Fig. \ref{Distribution_comp}). As noted in other works \cite{NT2}, electron-electron collisions primarily diffuse sharp peaks in the electron spectra. These collisions do not significantly affect the overall energy distribution, except for the thermal part, whose shape is highly influenced by elastic processes. In contrast, CI is the mechanism that transports electrons from the high-energy tail. It creates large plateaus between peaks as a result of many small energy collisions. This occurs because almost any hot electron has enough energy to ionize outer shells, and this energy is much smaller than the electron's total energy.

While modifying $\sigma_{CI-M}$ does not affect photo and auger peaks, it has an important effect on the thermalization. Figure \ref{Distribution_comp} shows how lower $\sigma_{CI-M}$, leads to bigger deviations from equilibrium, stacking more electrons at high energies. This effect is important since it delays the heating of the thermal portion of the distribution, which will affect the plasma population in time, especially for the M-satellites.

To quantify deviations from equilibrium, we define the \(\delta\) parameter, which indicates the percentage of electrons that are out of equilibrium:

\begin{equation}
    \delta = \frac{\int \mid f(\epsilon) - f_{\text{eq}}(\epsilon) \mid \, d\epsilon}{2 n(t)} \cdot 100,
\end{equation}  

\noindent where \(n(t)\) is the total electron density,  \(f(\epsilon)\) is the instantaneous electron distribution, and \(f_{\text{eq}}(\epsilon)\) is the equivalent equilibrium distribution, where the temperature and chemical potential are calculated by means of the total energy and density of the electronic system. It represents the percentage of electrons out of equilibrium. $\delta$ is most sensitive to deviations in the low-energy region of the electron distribution, as high-energy electrons, being orders of magnitude fewer in number, contribute minimally to the integral. Therefore, \(\delta\) serves as an effective measure of the thermalization delay.

\begin{figure}
    \centering
    \includegraphics[scale=0.7]{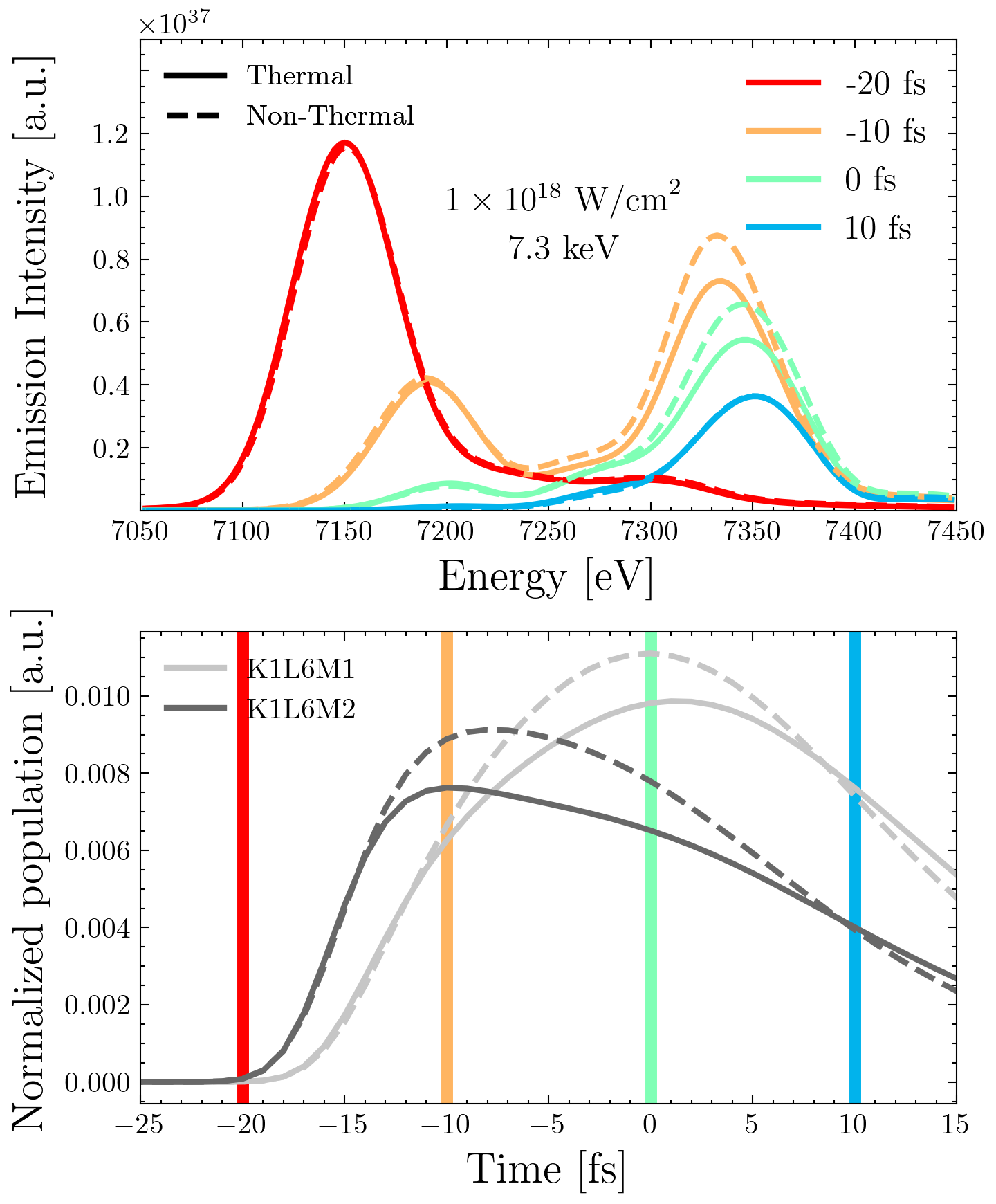}
    \caption{\label{ER}Comparison of thermal and non-thermal self-emission as a function of time (top) and the evolution of the most significant contributing configurations (bottom) for an intensity of \(1 \times 10^{18}\ \text{Wcm}^{-2}\), a 30 fs pulse duration, and a pump energy of 7.3 keV. Vertical color-coded lines in the bottom plot mark specific timestamps shown in the top plot. In the time axis, 0 corresponds to the peak of the pulse. 
}
\end{figure}

Figure \ref{delta} illustrates the evolution of the $\delta$ parameter for simulations with both regular and suppressed CI cross sections. It reveals two distinct regimes. The second one corresponds to overall thermalization, reaching its maximum around the peak of the pulse. In contrast, the first peak arises from non-thermal contributions in the low-energy region, driven by M-shell 3B processes in the full simulation, and by $LMM$-AU and $K$-shell photoionization in the elastic case. As the distribution heats up, elastic processes diffuse these low-energy contributions, resulting in the two observed behaviours. The findings clearly demonstrate that in the absence of CI, the system exhibits significantly larger deviations from equilibrium.

\begin{figure}
    \centering
    \includegraphics[scale=0.7]{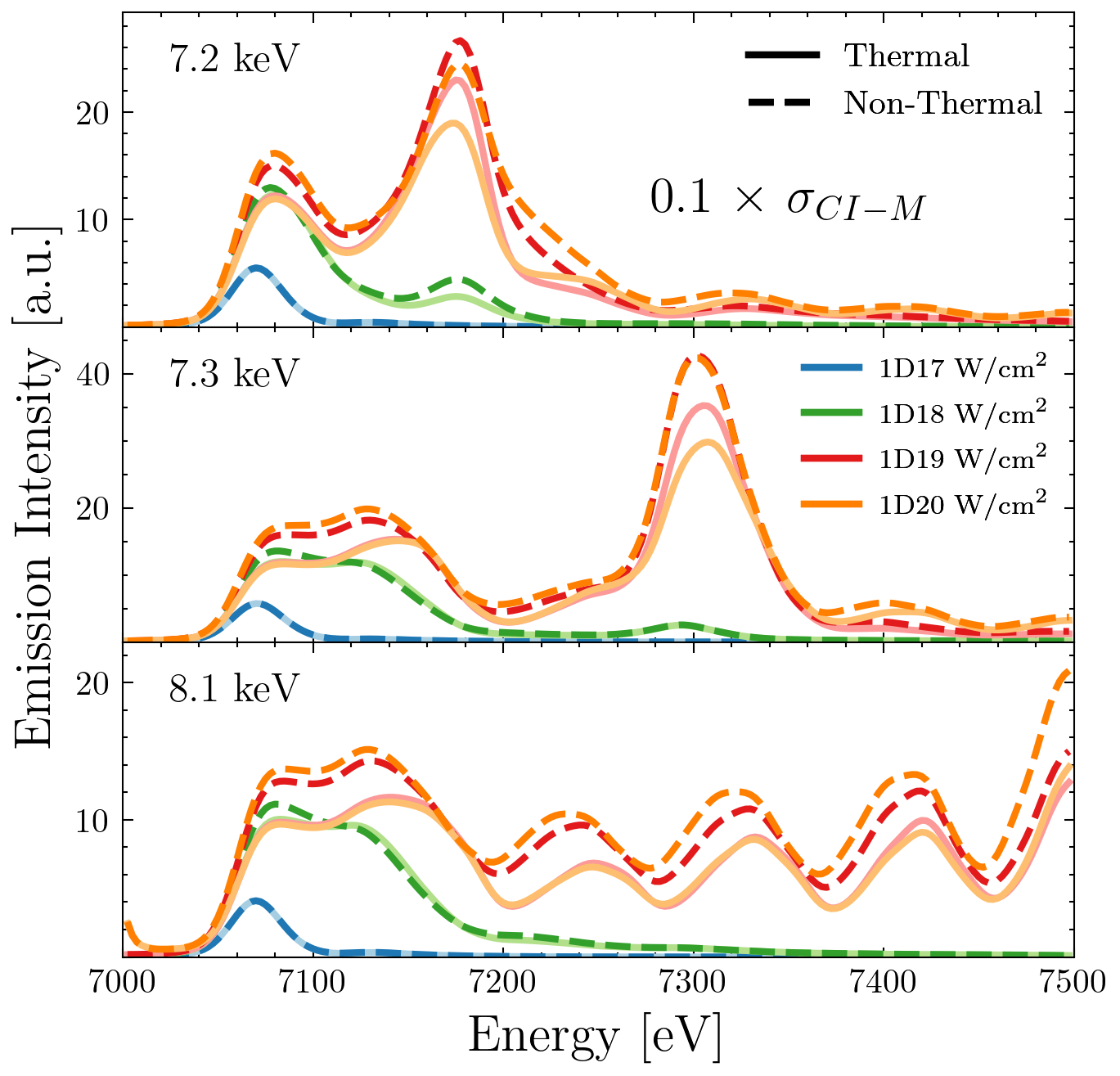}
    \caption{\label{S3RD}Comparison of thermal and non-thermal spectra at varying intensities for the three pump energies analysed, with the M-shell collisional ionization cross section reduced by a factor of 10. Results are shown for a 30 fs FWHM pulse.}
\end{figure}

Another feature, mainly important at low temperatures, is the collisional ionization paths only available for the high-energy non-equilibrium electrons. This is better represented in Figure \ref{Rates}, where the comparison of CI and 3B rates for instant thermalized and non-thermal simulations, together with the temporal evolution of the distribution function, is shown. At the beginning of the XFEL pulse, only hot electrons can produce CI because the thermal distribution is confined within the Fermi energy, which is typically lower than the threshold required to ionize even the outermost shell. It can be seen that only non-thermal features are above the M threshold in the first 30 fs, where $K$-shell PI and 3B are the most important sources. This is reflected in the rates, where the non-equilibrium simulation shows orders of magnitude higher CI. As the distribution heats up, rates converge from both simulations. This is a consequence of the thermalized distribution being able to ionize outer shell, which as have been already identified, is the main mechanism for the plasma evolution. As a consequence, both simulations will present relatively similar trends in overall ionization and complex evolution.

The case of 3B is slightly different. Because this process is the main electron capture mechanism, it tends to balance any source of electrons. For the thermal simulation, in the beginning, it just tends to compensate for PI and AU effects, while in the other case it closely follows CI, as it is already the most efficient ionization source.

\section{Spectral Signatures}\label{sec4}

High-energy-density plasma experiments rely heavily on spectral analysis as the primary diagnostic tool, especially since extremely short timescales, often on the order of femtoseconds, make other methods challenging to implement.

For iron and other $3d$ transition metals, K-shell spectra are divided into $K_{\alpha}$ and $K_{\beta}$ regions. The $K_{\alpha}$ region includes all possible $L$-to-$K$ transitions, while the $K_{\beta}$ region encompasses $M$-to-$K$. Both regions contain a combination of $L$ and $M$ satellite transitions; however, $M$ satellites in the $K_{\beta}$ region are more widely separated. This greater dispersion enhances the sensitivity to plasma evolution, making it easier to observe line shifts and broadening effects under varying conditions. Consequently, in this study, we focus on the first 4-5 $K_{\beta}$ $L$ satellites.

Each ionized $L$ electron emits radiation in an energy range of approximately 30 eV, depending on the number of $M$ electrons (from 0 to 8). In the $K_{\beta}$ region, these spectral bands often overlap, with the tail of one $L$ satellite region frequently coinciding with the beginning of the next (see, e.g.,  182 $\sim$ 177)  . Because multiple lines lie close to the pump energy, it becomes possible to effectively excite several of them, even if they are slightly off resonant. The efficiency and spectral position of these resonant lines are influenced by the overall plasma state.

In order to better understand the overall effect of non-thermal distributions, three different pump photon energies will be explored. In each setup, we scan a range of intensities and pulse durations. The first scenario uses a pump energy of 7.2 keV, positioned near the K-edge absorption, still allowing to ionize all $M$ electrons in the first $L$ satellite. The second uses a pump energy of 7.31 keV, aligning with the overlap between the second and third satellites. Finally, the third scenario involves a pump energy far from the initial satellites, allowing to investigate the effects of off-resonant pumping on satellite features.

In our analysis, we examine both non-thermal effects and the sensitivity to collisional ionization for the different XFEL photon energies. The XFEL parameters chosen are readily available at current facilities,providing a realistic basis for comparison. \cite{SMID, LEE}.

\subsection{Spectrum Under Non-Thermal Conditions}

The primary objective of this work is to determine whether non-thermal effects play a significant role and, if so, under which specific conditions they become prominent. In Fig. \ref{S3R}, we compare spectra across intensities ranging from \(1 \times 10^{17}\) to \(1 \times 10^{20}\ \text{Wcm}^{-2}\) for a pulse duration of 30 fs, that will corresponds to final thermalized temperatures ranging from 30 to 750 eV. Our results show trends similar to those observed in previous studies on lower-\( Z \) materials \cite{NT1}, where higher intensities enhance satellite line intensities under non-thermal electron conditions. At lower intensities, satellites are very weak, and the differences are minimal. There are no appreciable line shifts or new lines present in the non-equilibrium simulations. 

\begin{figure}
    \centering
    \includegraphics[scale=0.7]{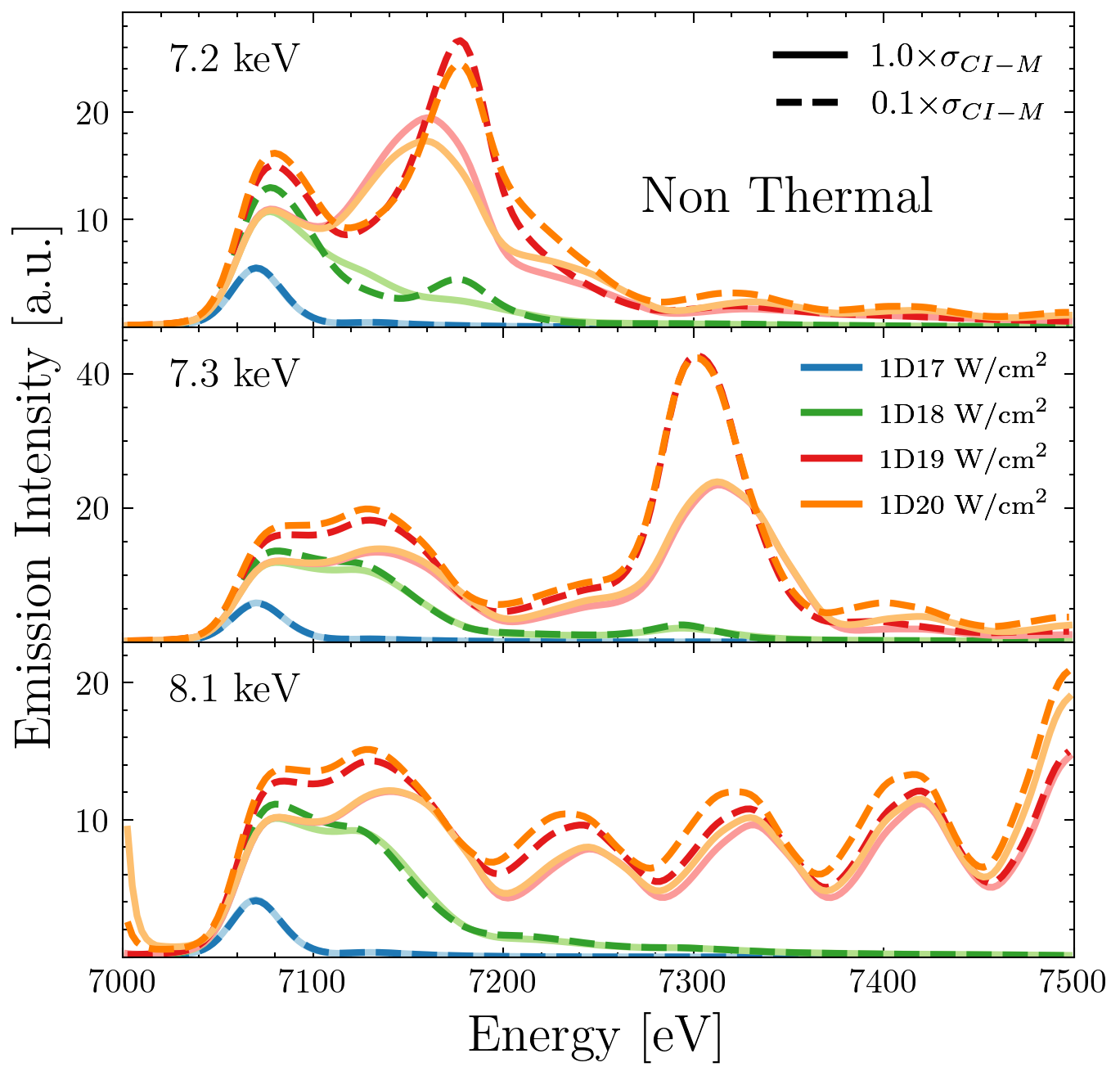}
    \caption{\label{S3RRD}Comparison of non-thermal simulation spectra with regular and reduced cross sections, plotted as a function of intensity for the three pump energies discussed. Results are shown for a 30 fs FWHM pulse.}
\end{figure}

To further investigate the underlying mechanisms, we analyse the evolution of the radiation emission. In particular, focusing on the case of 7.3 keV, we track the super-configurations 162 and 161, which are among the strongest contributors to the resonant line. In this case, satellites are mainly generated either through resonant pumping or by imbalances between CI and 3B processes, as the plasma approaches non-local thermodynamic equilibrium (NLTE).  

In Fig. \ref{ER} (top) is represented the emitted spectrum at different simulation times, while the bottom plot shows the evolution of the specified super-configurations within the $L6$ satellite. Not appreciable delay is found in the complex creation, but the non-thermal simulation shows a faster growth rate. This occurs because, with a non-thermal distribution, more electrons reside above the ionization threshold, leading to a greater CI.  Although there are apparent differences in ionization rates between thermal and non-thermal conditions (reflected in the time-resolved emission and ionization rates shown in Fig. \ref{S3R}) these variations result in only minor differences in the time-averaged spectra, as depicted in Fig. \ref{ER}.

%Pulse duration has a similar impact, as shorter FWHM for the same energy results in higher peak intensity, enhancing non-thermal effects. APENDIX?

\subsection{Cross-Section Influence}

\begin{figure}
    \includegraphics[scale=0.7]{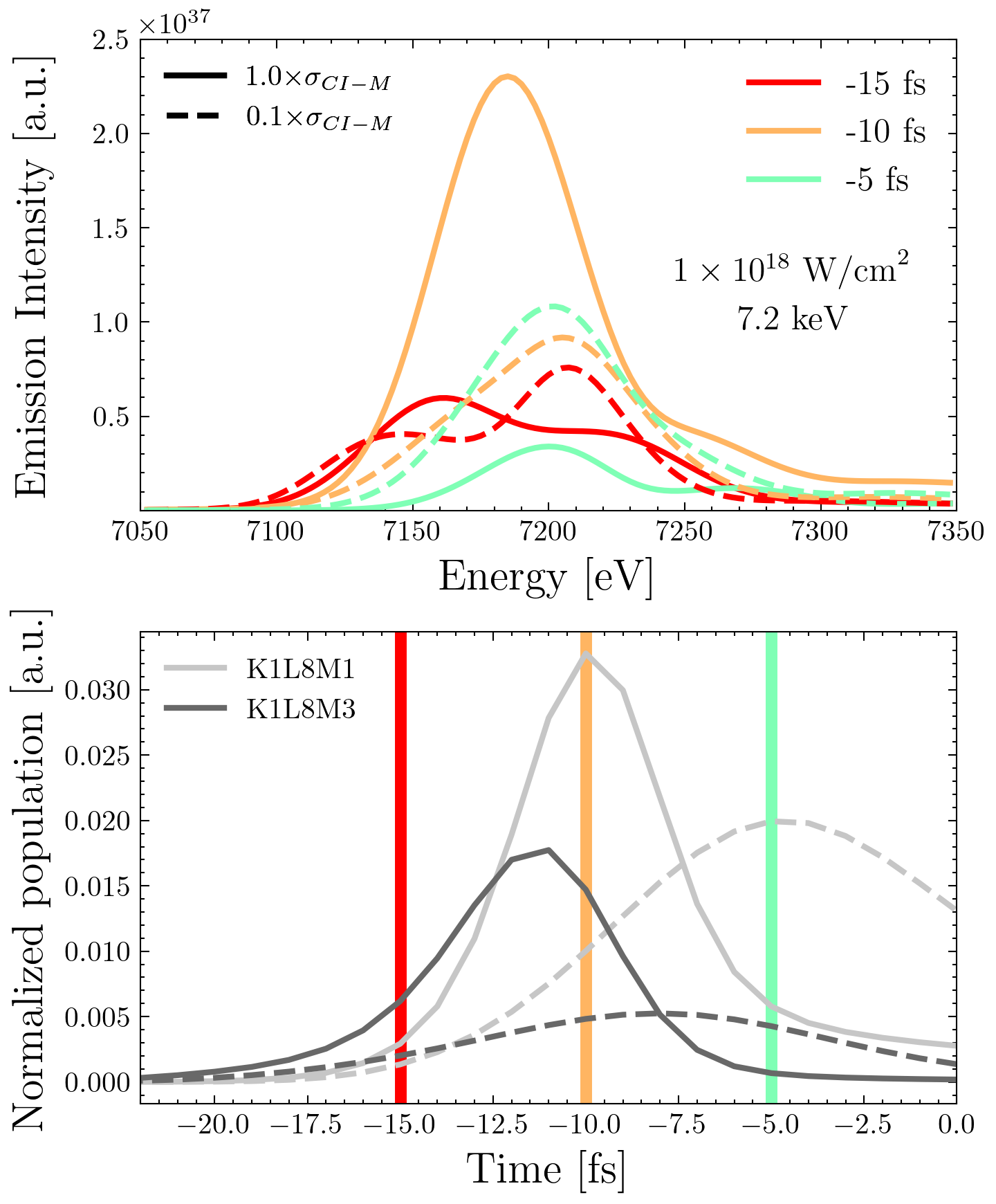}
    \caption{\label{ERRD}Comparison of time-dependent emission for reduced and regular cross sections as a function of time (top) and the evolution of the most significant contributing configurations (bottom) for an intensity of \(1 \times 10^{18}\ \text{Wcm}^{-2}\), a 30 fs pulse duration, and a pump energy of 7.2 keV. Vertical color-coded lines in the bottom plot denote specific timestamps shown in the top plot. In the time axis, 0 corresponds to the peak of the pulse.
}
\end{figure}

Reducing collisional ionization cross sections can enhance non-thermal effects because thermalization times are influenced by inelastic collisions. To see how this affects self-emission, we revisited the earlier cases with simulations where the $\sigma_{CI-M}$ values were reduced by a factor of 10 (Fig \ref{S3RD}).

 The emission spectra with a reduced $M$-shell cross- section in Fig. \ref{S3RD} show a larger difference between thermal and non-thermal cases at high intensities (\(1 \times 10^{19}\), \(1 \times 10^{20}\ \text{Wcm}^{-2}\)) than the standard $M$-shell cross sections in Fig 4. In addition to the previously observed satellite enhancements, the main $K_{\beta}$ line also shows increased intensity. Additionally, satellite shifts become noticeable at higher intensities, especially in the off-resonance spectra, where a general blue shift of 6–7 eV is observed. These shifts arise from differences in the relatives $M$-satellite populations during emission.

\begin{figure*}
    \includegraphics[scale=1]{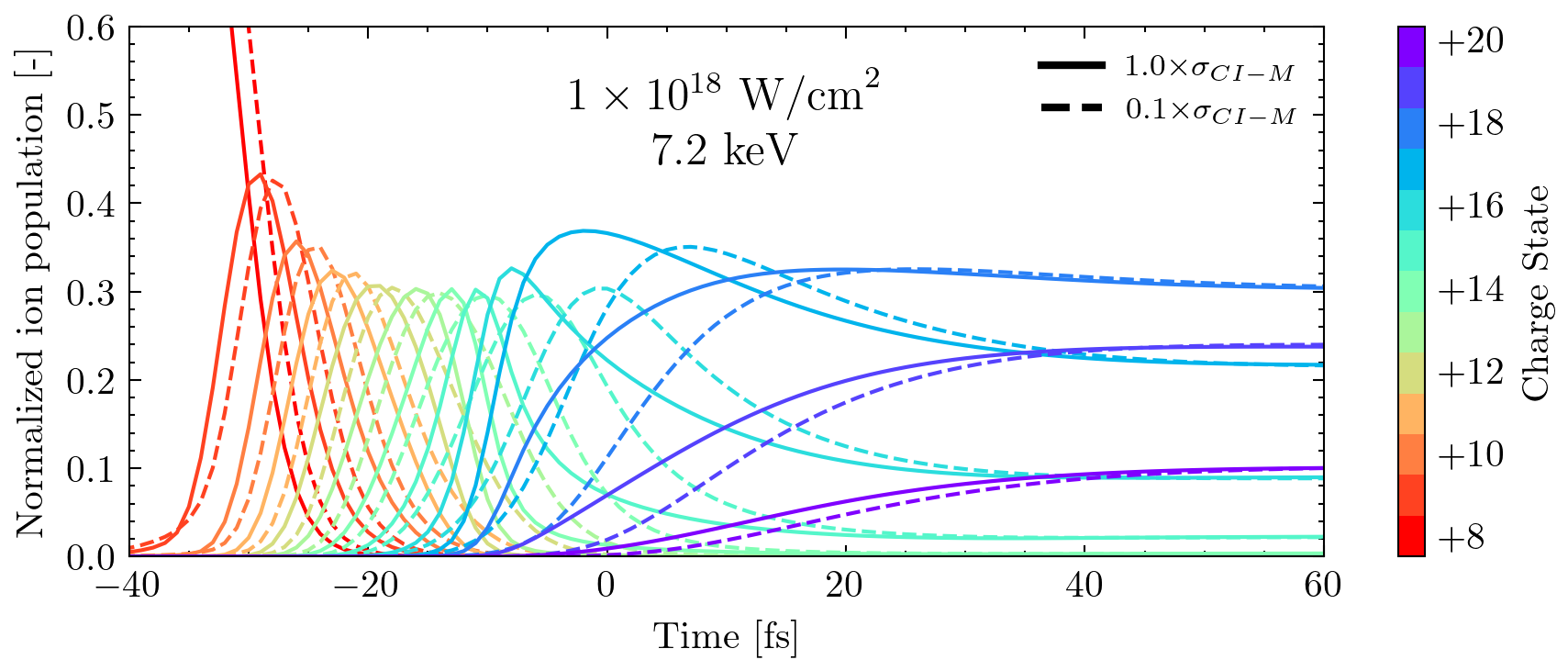}
    \caption{\label{COMRD} Comparison of the evolution of ionic states as a function of time for reduced and regular cross sections, with an intensity of \(1 \times 10^{18}\ \text{Wcm}^{-2}\), a 30 fs pulse duration, and a pump energy of 7.2 keV. In the time axis, 0 corresponds to the peak of the pulse.}
\end{figure*}

This effect becomes more evident when comparing the emission from two non-thermal simulations: one with regular cross sections and the other with reduced cross sections. The results are shown in Figure \ref{S3RRD}. Shifts are now observed for all pumping energies. To further understand the origin of this signature, we track the emission and evolution of the 181 and 183 complexes, which are the main contributors to the 7.2 keV resonant satellite.

Figure \ref{ERRD} shows the complex emission comparison for the 8.1 keV pumping case. The differences in this case arise not only from the reduced cross section allowing a build-up of more non-thermal electrons but also from the altered timescales in the overall plasma evolution. In the regular simulation, the intensities of the $M1$ and $M2$ satellites are nearly equal at the peak. However, in the reduced cross-section case, the $M1$ satellite becomes significantly stronger. This disparity ultimately leads to the observed blue shift in the emission spectrum. Resonant lines are also significantly more pronounced. This arises from two main factors. First, the satellites have a longer lifetime due to the overall delay. Second, these satellites appear closer to the pulse peak, resulting in more efficient resonant pumping.  

Although these simulations show significant differences during their evolution, it is important to note that the cross sections were modified according to Fowler’s relation. As a result, the final equilibrium conditions are the same for both cases (Fig. \ref{COMRD}). In other words, the energy deposition remains consistent, leading to identical final states despite the distinct temporal behaviours observed.

\section{Conclusions}\label{conc}

In this study, we investigated the impact of non-equilibrium electron distributions on the evolution and spectral signatures of isochorically heated solid-density iron. We identified collisional ionization of outer $M$-shell electrons as the primary thermalization mechanism, leading to the formation of non-equilibrium plateaus between resonant peaks in the electron distribution.

Non-thermal electrons influence the plasma evolution, particularly during the initial femtoseconds, where enhanced collisional ionization rates from high-energy electrons as the dominant collisional pathway. As the plasma evolves, these rates gradually converge with those of the thermalized distribution. Additionally, the formation of the thermalized electrons can be delayed under non-equilibrium conditions, slowing overall plasma relaxation.

These changes in the plasma's temporal evolution are directly reflected in the emitted spectra. Distinct signatures in the $K_{\beta}$ spectra were observed at higher intensities that resulted in stronger satellite lines. These effects were further enhanced when collisional cross sections were reduced, as expected due to slower thermalization and an increased number of electrons driving L-shell CI processes, resulting in even more prominent satellites.     

Additionally, the $K_{\beta}$ spectra were found to be sensitive to variations in $\sigma_{CI-M}$, with 5 - 7 eV line shifts occurring under different cross-section assumptions, that would be experimentally observable. This sensitivity provides a potential avenue for refining collisional cross-section models using either existing experimental data or new campaigns at XFEL facilities, where the simulated conditions are already achievable.

\begin{acknowledgments}

L. A. acknowledge support from by FCT (Foundation for Science and Technology - Portugal) under Grant No. UI/BD/153734/2022. L. A., M. F. and G. W. also acknowledge FCT support under the project UIDB/50010/2020. P. V. and L. A. acknowledge the support of the Spanish Government through the project PID2021-124129OB-I00 funded by MCIN/AEI/10.13039/501100011033/ERDF. G. W., This work was supported by FCT Concurso de Projetos de ID em Todos os Dominios Cientificos: 2022.09213.PTDC. FCT. and Concurso Esttímulo ao Emprego Científico Individual CEECIND (DOI:10.54499/2022.00804.CEECIND/CP1713/CT0003).

This project has received funding from the European Union's Horizon Europe programme, European Research Council EIC Pathfinder Open under grant agreement No 101047223 (NanoXcan).

The authors gratefully acknowledge the Universidad Politécnica de Madrid (www.upm.es) for providing computing resources on Magerit Supercomputer.

\end{acknowledgments}

%\nocite{*}

\bibliographystyle{apsrev4-2}
\bibliography{apssamp}% Produces the bibliography via BibTeX.

\end{document}